\documentclass[final]{svjour2}
\usepackage{rotating}
\usepackage{amssymb}
\usepackage{mathptmx}
\usepackage{graphicx}
\usepackage{array}
\usepackage{ marvosym }
\usepackage{enumitem}
\usepackage{rotating}
\usepackage[numbers]{natbib}
\makeatletter
\journalname{Journal of Low Temperature Physics}

\bibpunct{}{}{,}{s}{}{,}

\begin{document}

\newcommand{\hdblarrow}{H\makebox[0.9ex][l]{$\downdownarrows$}-}
\title{A Study of Al-Mn Transition Edge Sensor Engineering for Stability}

\author{E. M. George$^{1}$ \and J. E. Austermann$^{3}$ \and J. A. Beall$^{4}$ \and D. Becker$^{4}$ \and B. A. Benson$^{2,13}$ \and L. E. Bleem$^{2,12}$ \and J. E. Carlstrom$^{2,5,12,13,14}$ \and C. L. Chang$^{2,5,13}$  \and H- M. Cho$^{4}$ \and A. T. Crites$^{2,14}$ \and M. A. Dobbs$^{6}$ \and W. Everett$^{3}$ \and N. W. Halverson$^{3,11}$ \and J. W. Henning$^{3}$ \and G. C. Hilton$^{4}$ \and W. L. Holzapfel$^{1}$ \and J. Hubmayr$^{4}$ \and K. D. Irwin$^{4}$ \and D. Li$^{4}$ \and M. Lueker$^{8}$ \and J. J. McMahon$^{9}$ \and J. Mehl$^{2,13}$ \and J. Montgomery$^{2,12}$ \and T. Natoli$^{2,12}$ \and J. P. Nibarger$^{4}$ \and M. D. Niemack$^{4}$ \and V. Novosad$^{7}$ \and J. E. Ruhl$^{10}$ \and J. T. Sayre$^{10}$ \and E. Shirokoff$^{8}$ \and K. T. Story$^{2,12}$ \and G. Wang$^{5}$ \and V. Yefremenko$^{5,7}$ \and K. W. Yoon$^{4}$ \and E. Young$^{1}$}

\institute{
\email{lizinvt@berkeley.edu} (E. M. George)\\
$^{1}$ University of California, Berkeley, 151 LeConte Hall Berkeley, CA 94720, USA\\
$^{2}$ Kavli Institute for Cosmological Physics, Department of Physics, Enrico Fermi Institute, The University of Chicago, Chicago, IL 60637, USA\\
$^{3}$ Department of Astrophysical and Planetary Sciences, University of Colorado, Boulder, Colorado,80309, USA\\
$^{4}$ NIST, Boulder, CO 80305, USA\\
$^{5}$ High Energy Physics Division, Argonne National Laboratory, Argonne, IL 60439, USA\\
$^{6}$ McGill University, Montreal, Quebec, Canada\\
$^{7}$ Materials Science Division, Argonne National Laboratory, Argonne, IL 60439, USA\\
$^{8}$ California Institute of Technology, Pasadena, CA 91125, USA\\
$^{9}$ University of Michigan, Ann Arbor, Michigan, USA\\
$^{10}$ Case Western Reserve University, Cleveland, Ohio 44106, USA\\
$^{11}$ Department of Physics, University of Colorado, Boulder, CO 80309\\
$^{12}$ Department of Physics, University of Chicago, 5640 South Ellis Avenue, Chicago, IL, USA 60637 \\
$^{13}$ Enrico Fermi Institute, University of Chicago, 5640 South Ellis Avenue, Chicago, IL, USA 60637\\
$^{14}$ Department of Astronomy and Astrophysics, University of Chicago, 5640 South Ellis Avenue, Chicago, IL, USA 60637
}

\date{\today}

\maketitle

\begin{abstract}
The stability of Al-Mn transition edge sensor (TES) bolometers is studied as we vary the engineered TES transition, heat capacity, and/or coupling between the heat capacity and TES. We present thermal structure measurements of each of the 39 designs tested. The data is accurately fit by a two-body bolometer model, which allows us to extract the basic TES parameters that affect device stability. We conclude that parameters affecting device stability can be engineered for optimal device operation, and present the model parameters extracted for the different TES designs.


\end{abstract}

\section{Introduction}

High-sensitivity measurements of the cosmic microwave background (CMB) polarization can constrain the sum of the neutrino masses and the energy scale of inflation, which informs the viability of inflationary models. To make these measurements, we developed 84-pixel arrays of 150 GHz Al-Mn transition edge sensor (TES) polarimeters\cite{henning2012} for the South Pole Telescope polarimeter (SPTpol), which began observations in February 2012 \cite{george2012}. The detectors are read out with a digital frequency domain multiplexing (fMUX) system.\cite{dobbs2008} Initial detector prototypes exhibited instability consistent with a compound TES model, described in \citet{lueker2008} when operated at moderate depths in the superconducting transition.\cite{hubmayr2010} In these proceedings, we describe a study of 40 different device designs that were devised to address TES stability criteria.

\section{Stability Criteria for TESes}

\begin{figure}
\begin{center}
\includegraphics[
  width=0.7\linewidth,
  keepaspectratio]{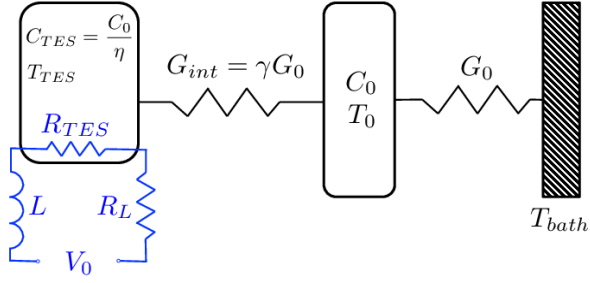}
\end{center}
\caption{(Color online) Black: Two-body thermal model used to model the thermal structure of our bolometers. The node in the middle represents the BLING with heat capacity $C_{0}$, coupled to the bath with a thermal link $G_{0}$. The left node is the TES, which is strongly coupled to the BLING with a thermal link $G_{int} = \gamma G_{0}$. Blue: Electrical circuit for the TES, with inductance $L$ and parasitic resistance $R_L$.}
\label{fig:2body}
\end{figure}

Figure \ref{fig:2body} is a model of a TES with an additional heat capacity, commonly known as a ``Bandwidth Limiting Interface Normally of Gold'' (BLING), coupled to the TES through a thermal link. The TES has a heat capacity $C_{TES} = C_0/\eta$, resistance $R_{TES}$, and is strongly coupled to the BLING by a thermal conductance $G_{int} = \gamma G_0$. The BLING has a heat capacity $C_0$, which is connected to the thermal bath by a conductance $G_0$. In our devices, the BLING is strongly coupled to the TES, ($\gamma \gg 1$) and the heat capacity of the BLING is much greater than the heat capacity of the TES, ($\eta \gg 1$). The TES is AC voltage biased in negative electrothermal feedback (ETF) with loopgain $\mathcal{L} = \frac{\alpha P_{e}}{G_{0} T_{c}}$ , where $\alpha$ is the logarithmic derivative of resistance with temperature, $\alpha = \frac{\partial ln(R_{TES})}{\partial ln(T)}$, $P_{e}$ is the electrical bias power, and $T_c$ is the superconducting transition temperature. The TES is read out with fMUX readout with an electrical time constant of $\tau_e = 2L/R_{TES}$. 

The bandwidth limit of the readout imposes a stability requirement on the thermal time constant of the detector ($\tau_{th}$) of:
\begin{equation}
\tau_{th} = \frac{C_{0}}{G_{0}}\frac{1}{\mathcal{L}+1} > 5.8\tau_{e}.
\label{eq:constraint1}
\end{equation}
Equation \ref{eq:constraint1} is the extension of the one-body (simple TES) stability criteria to a two-body device in the limit that $\gamma \rightarrow \infty$ and $\eta \rightarrow \infty$. The original criteria is derived in \citet{irwin1998} and \citet{irwin2005} by requiring that the eigenvalues of the responsivity matrix be negative and real-valued. Equation \ref{eq:constraint1} is the same as the stability criterion for a one-body TES, except $C_{TES}$ is replaced by $C_0$ here. 

In a real compound device, the BLING is not perfectly coupled to the TES, and the TES has a finite heat capacity: $\gamma, \eta \gg1$, but neither is infinite. The responsivity matrix is now more complicated (see \citet{lueker2011}) and the stability requirement that comes from requiring negative and real valued eigenvalues in the limit that $\tau_{th} \gg \tau_e$ (i.e. the limit that equation \ref{eq:constraint1} has already been satisfied) is: 
\begin{equation}
\mathcal{L} < \gamma + 1 + \frac{C_{TES}}{G_{0}\frac{\gamma}{\gamma + 1} \tau_{e}}.
\label{eq:constraint2}
\end{equation}
If $C_{TES}$ is sufficiently large, then the device can remain stable without additional heat capacity. In our devices, $C_{TES}$ is much too small and an additional heat capacity $C_0$ must be added to satisfy equation 1. Because of this, the last term in equation \ref{eq:constraint2} can be ignored, and equation \ref{eq:constraint2} becomes a constraint on $\gamma$.

To meet these two stability criteria we need to engineer our bolometers to have a higher $\tau_{th}$ and a higher $\gamma$ (and $G_{int}$). We can increase the thermal time constant by adding heat capacity ($C_0$) to the TES island\cite{lueker2008}$^{,}$\cite{shirokoff2009}, or by decreasing the loopgain ($\mathcal{L}$). $G_0$, $T_c$, and $P_{e}$ are constrained by observational requirements, so we can only decrease $\mathcal{L}$ by decreasing $\alpha$, which can be accomplished with the addition of normal metal structures on the TES.\cite{staguhn2004} $\gamma$ can be increased by improving the interface between the TES and BLING.

\section{Device designs}\label{sec:designs}

The base TES design is a 45 nm thick, 48 $\mu$m wide Al-Mn TES with $\sim$1 $\Omega$ normal resistance ($R_n$) and a transition temperature of $\sim 540$ mK. The TES has niobium leads overlapping the TES material on the two ends. Four nitride legs support the TES and provide the thermal link to the bath. The thermal conductance of the legs is $G_{0} \sim 120$ pW/K, to achieve a saturation power of $\sim 22$ pW at a bath temperature of $280$ mK. To make devices with lower $\alpha$ (and $\mathcal{L}$), higher $C_0$, and higher $G_{int}$ (and $\gamma$), we considered four types of modifications to our basic design:
\begin{enumerate}
\item Addition of gold BLING of various thicknesses to increase heat capacity ($C_0$) and a range of geometries which probe the thermal coupling between the TES and BLING ($\gamma$).
\item Addition of normal metal features (bars or dots) on the TES to soften the TES transition ($\alpha$) and consequently decrease device loop gain ($\mathcal{L}$).
\item Substituting PdAu (52\% Pd by mass, DC sputter from an alloy target) for the Au BLING to decrease the thickness of BLING needed for a given heat capacity ($C_0$).
\item Addition of a gold cap on top of an insulator above the TES to improve TES-BLING coupling ($\gamma$). 
\end{enumerate}
Combinations and variations of these four basic design modifications led to 40 different device designs. The resulting devices are listed in Table \ref{tab:TESparams}. All devices were fabricated on a single wafer with 10 devices per die (1 cm x 1 cm) and 4 different types of dies corresponding to the devices in each column in Table \ref{tab:TESparams}. This ensured uniformity of the basic TES parameters such as normal resistance ($R_n$) and thermal conductance ($G_0$), allowing us to directly compare design changes. In the case of uniform $G_0$, $\gamma$ can be used directly as a proxy for $G_{int}$. Accounting for systematics in our test setup, these parameters were measured to be uniform at the $\sim5\%$ level. 

\section{Measurements and Analysis}\label{sec:anal}

To rapidly evaluate these 40 designs we used a simple technique described in \citet{lueker2008} and \citet{lueker2011} to measure the internal thermal structure of these devices using frequency multiplexed readout. A TES is voltage biased with a carrier signal at a frequency $\omega_0$ and amplitude $V_0$. In addition to the carrier, we inject a small sinusoidal probe signal with amplitude $V^{\prime}$at a frequency $\omega_0 - \omega$ which will perturb the TES with a power $\delta P(\omega) = (V_0 V^{\prime})/(R_{TES} \sqrt{1 + \omega^2 \tau_e^2})$. The amplitude of the current measured in the opposite sideband, $\left| I_{sb}(\omega_0 + \omega)\right|$, is proportional to the power-to-current responsivity $s_i(\omega)$. Ignoring parameters that are expected to be negligible under our operating conditions ($\beta = \frac{\partial ln(R_{TES})}{\partial ln(I)} \ll 1$, $R_L\ll1$), the equation is $\left| I_{sb}(\omega_0 + \omega)\right| = \frac{V^{\prime}V_0}{R_{TES}}\frac{\left|s_i(\omega)\right|}{\left|1+i\omega\tau_e\right|}$, which expands to: 
\begin{equation}
\left|I_{sb}(\omega_0+\omega)\right| = 
\frac{V^{\prime} \mathcal{L}}{R_{TES} \sqrt{1+\omega^2 \tau_e^2}} \left|\left(\frac{G(\omega)}{G_{eff}}(1+i\omega \tau_e) + \mathcal{L}(1-i\omega \tau_e)\right)^{-1}\right|
\label{eq:Iresp}
\end{equation}
where $G(\omega)$ is the generalized thermal conductance defined in \citet{lanting2005} and $G_{eff}$ is the effective thermal conductance at the TES, which for the two-body model is $G_{eff} = G_0\frac{\gamma}{\gamma+1}$.

The form of $G(\omega)$ depends on the bolometer thermal model chosen, and for the two-body model is
\begin{equation}
G(\omega) = G_0\frac{\gamma}{1+\gamma}\left(\frac{1+i\omega \frac{C_0}{G_0}}{1+i\omega\frac{C_0}{(1+\gamma)G_0}}\right).
\label{eq:GofW}
\end{equation}
Equation \ref{eq:GofW} can be combined with equation \ref{eq:Iresp} to obtain a model for $|I_{sb}(\omega_0 + \omega)|$ which contains only relevant device parameters ($R_{TES}$, $\tau_0 = C_0/G_0$, $\gamma = G_{int}/G_0$, and $\mathcal{L}$) and parameters of the system ($\tau_e$ and $V^{\prime}$). $\tau_0$ is the intrinsic thermal time constant of the detector as $\eta, \gamma \rightarrow \infty$ and $\mathcal{L} \rightarrow 0$.

We bias each TES to a depth in the superconducting transition, $f_{R}=R_{TES}/R_n$, and measure $|I_{sb}(\omega_0 + \omega)|$ at probe offset frequencies ($\omega/2\pi$) from 3-40,000 Hz, and repeat this measurement for $f_R$ from 0.6-0.99. We extract $\tau_0$ by fitting the data trace taken at $f_R$=0.99 to equations \ref{eq:Iresp} and \ref{eq:GofW} using a fixed, low loop gain ($\mathcal{L} \sim 0$). We then simultaneously fit all remaining data traces to equations \ref{eq:Iresp} and \ref{eq:GofW}, fixing $\tau_0$ to the value extracted from the fit at $f_R$=0.99. To simplify modeling $\gamma$ is constrained to be the same at each $f_R$. $\tau_e$ is fixed individually in each trace to be $2L/(f_{R} R_n)$. We allow $\mathcal{L}$ to vary in each trace, denoted as $\mathcal{L}(f_{R})$. 

\begin{figure}
\begin{center}
\includegraphics[
  width=0.75\linewidth,
  keepaspectratio]{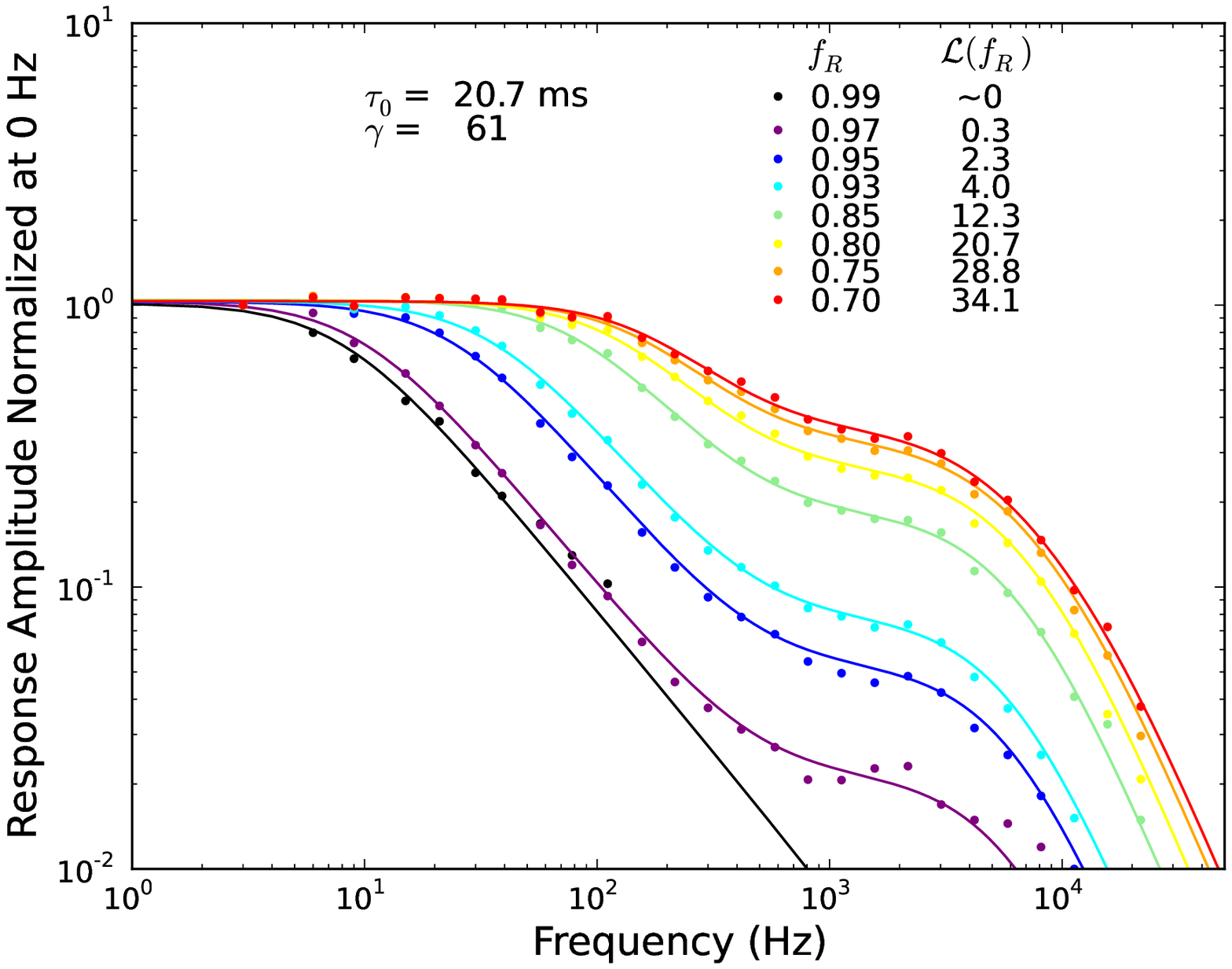}
  \includegraphics[
  width=0.75\linewidth,
  keepaspectratio]{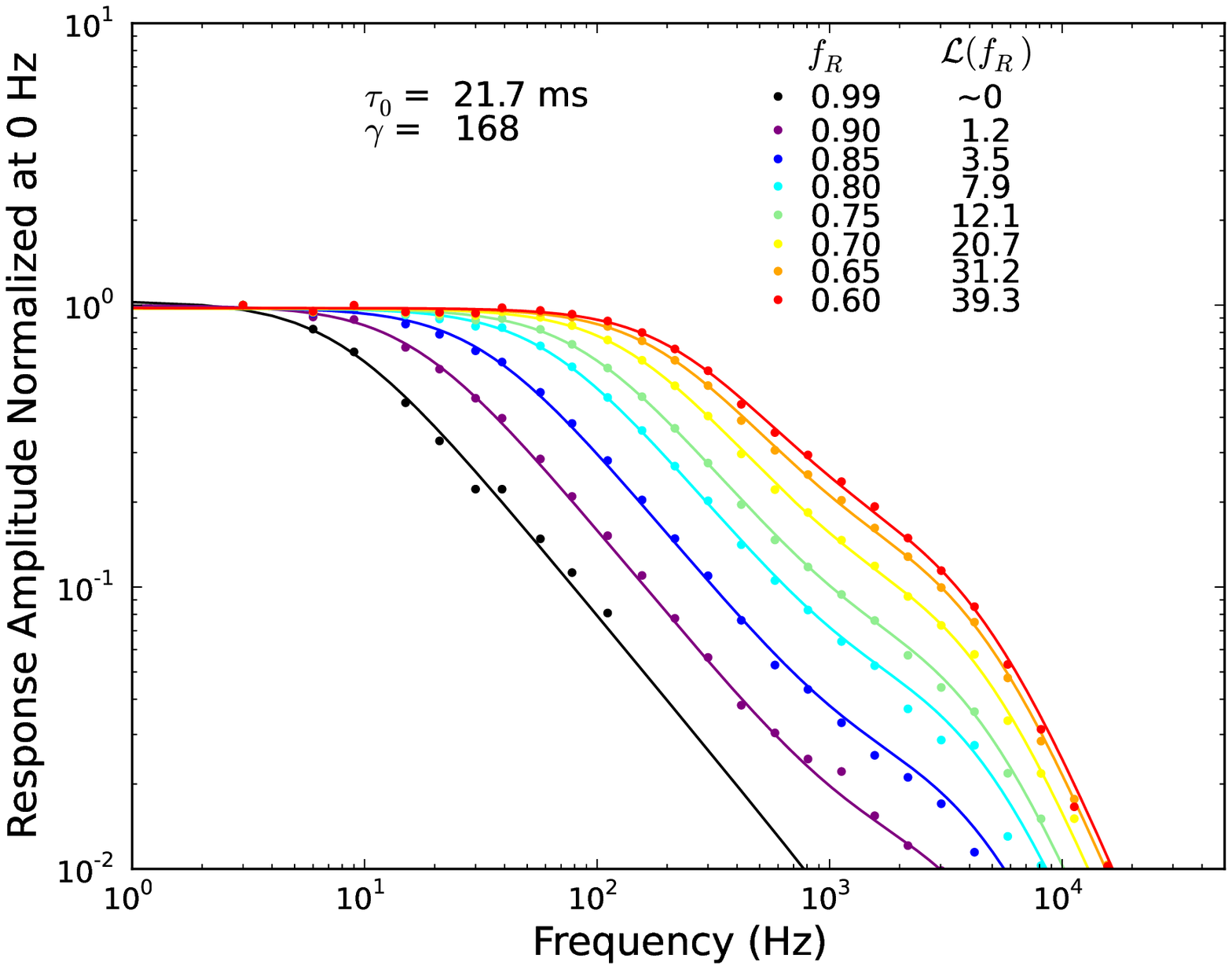}
\end{center}
\caption{(Color online) Normalized frequency response of two detectors with similar intrinsic time constants ($\tau_0$) but vastly different internal coupling of the bling to the TES ($\gamma$). The dots are the measured data points and the lines are a fit to equations \ref{eq:Iresp} and \ref{eq:GofW}. Traces from left to right on the plot go from higher to lower $f_R= R_{TES}/R_n$. $\mathcal{L}(f_{R})$ is the loop gain at $f_{R}$. The models were fit using the method described in Section \ref{sec:anal}. In the device with lower $\gamma$ (top), the BLING decoupling is obvious at a frequency of $\sim$800Hz. Top: One detector of type 1,2. Bottom: One detector of type 1,5.}
\label{fig:dataplots}
\end{figure}

Of the 40 types of devices fabricated, 39 were measured and the resulting data fit to the two-body bolometer model using the procedure described in this section. Figure \ref{fig:dataplots} shows the data and model fits obtained for two TESes. 

\section{Discussion}

The two-body model described by equations \ref{eq:Iresp} and \ref{eq:GofW} adequately models the detector response over the range of the superconducting transition where we normally operate our detectors ($f_{R}$ = 0.6-0.99), and we can extract the parameters $\tau_0$, $\gamma$, and $\mathcal{L}(f_{R})$ from the fits to this model. Table \ref{tab:TESparams} lists the parameters extracted from the fits for each TES. The model parameters reported in Table \ref{tab:TESparams} can be used to evaluate the effectiveness of various design modifications described in Section \ref{sec:designs}. Comparing devices across a row or down a column in Table \ref{tab:TESparams} is a direct comparison of individual design changes.

Each row has a fixed TES-BLING interface, which allows comparison between the effects of adding different structures to the TES. For example, comparing across row 1 reveals that the addition of bars between type 1,1 and type 2,1 decreased $\mathcal{L}(0.6)$ by reducing the $\alpha$ of the superconducting transition, while not affecting the intrinsic time constant ($\tau_0$) or the TES-BLING coupling ($\gamma$). Comparing type 2,1 with type 4,1 reveals that the addition of a gold/insulator cap over the TES both decreases $\mathcal{L}(0.6)$ and increases $\gamma$ while leaving $\tau_0$ almost unchanged. Comparing devices down a column reveals the effect of changes to the TES-BLING interface geometry. For example, if we compare type 1,1 and type 1,5 in column 1, we find that simply extending the BLING past the Nb leads into the TES region increases the TES-BLING coupling ($\gamma$) by a factor of $\sim$2.

Evaluating the data on the group of devices as a whole, several trends can be seen. These are:

\begin{enumerate}
\item{The geometry of the TES-BLING interface is important for TES-BLING coupling. In particular, direct metal contact between the BLING and TES drastically increases $\gamma$, when compared to contact through an intermediate dielectric or superconducting barrier with similar physical dimensions.}
\item{The addition of bars or other structures on the TES lowers $\alpha$ (and $\mathcal{L}$).}
\item{The addition of a gold/insulator cap over the TES lowers $\alpha$ (and $\mathcal{L}$), and increases $\gamma$.}
\end{enumerate}

\section{Conclusions}

Using the two-body bolometer model to describe the thermal response of our TES samples, we extract model parameters that affect TES stability: $\tau_0$, $\mathcal{L}(f_{R})$ (and $\alpha$), and $\gamma$ (and $G_{int}$), for each of our various TES designs. By comparing the model parameters for each design, we can evaluate which TES design changes to employ to optimize our device operation. We find that various interfaces between the BLING and TES improve the BLING-TES coupling by factors of 2-3 with a $G_{int}$ ranging from $\sim$7-20 nW/K over the various designs. We also find that various structures on the TES can degrade $\alpha$, and hence $\mathcal{L}$, by factors of 2-8 at 0.6$R_n$, the deepest point in the transition that we typically operate our detectors. 

Our study of these 40 different TES samples resulted in 40 TES designs that could be operated stably at moderate loopgains, and showed none of the excess noise from poorly coupled BLING that early prototypes displayed (see section 5.2 of \citet{lueker2011}). The results of this study were incorporated into the design of stable TES detectors deployed in the SPTpol array.\cite{george2012} The fielded devices have 540 nm thick PdAu BLING, with a TES-BLING interface such that the BLING extends past Nb leads into TES region, and no bars or structures on the surface of the TES.

\begin{sidewaystable}[h!]
\vspace{8cm}
  \centering
  \begin{tabular}{|m{.35cm} | m{.7cm} m{3cm} m{.35cm} | m{.7cm}  m{3.5cm}  m{.35cm} | m{.7cm} m{3.5cm} m{.35cm} | m{.7cm} m{4cm} m{.35cm} | m{.05cm}}
    \cline{1-13}
    Var.
    &

    & 
    Type 1
    &  
     \begin{minipage}{8cm}
     	\vspace{.1cm}
        $\tau_{0}$ \\
        $\gamma$ \\
        $\mathcal{L^*}$
        \vspace{.1cm}
      \end{minipage} 
             &
      & 
      Type 2
      &  
      \begin{minipage}{5cm}
     	\vspace{.1cm}
        $\tau_{0}$ \\
        $\gamma$ \\
        $\mathcal{L^*}$
        \vspace{.1cm}
       \end{minipage} 
             &
             
      & 
      Type 3 
      &  
      \begin{minipage}{5cm}
     	\vspace{.1cm}
        $\tau_{0}$ \\
        $\gamma$ \\
        $\mathcal{L^*}$
        \vspace{.1cm}
       \end{minipage} 
             &
      & 
      Type 4
      &  
      \begin{minipage}{5cm}
     	\vspace{.1cm}
        $\tau_{0}$ \\
        $\gamma$ \\
        $\mathcal{L^*}$
        \vspace{.1cm}
       \end{minipage} 
       &
      \\ \cline{1-13}
      1
      &
    \begin{minipage}{.75cm}
      \includegraphics[angle=90,width=\linewidth,]{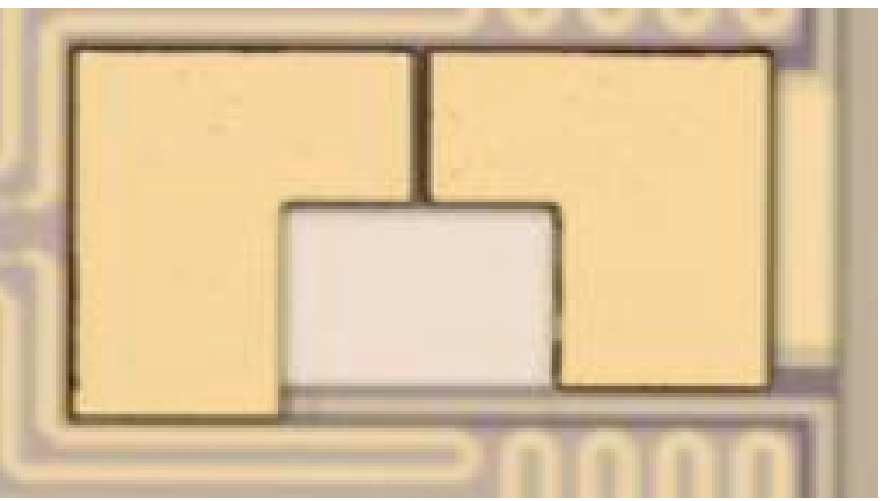}
    \end{minipage}
    &
	Symmetric Au bling 3350 nm. Usual interface.
    & 
    \begin{minipage}{5cm}
      24 \\ 
      84 \\ 
      51
    \end{minipage}
    &
     \begin{minipage}{.75cm}
      \includegraphics[angle=90, width=\linewidth,]{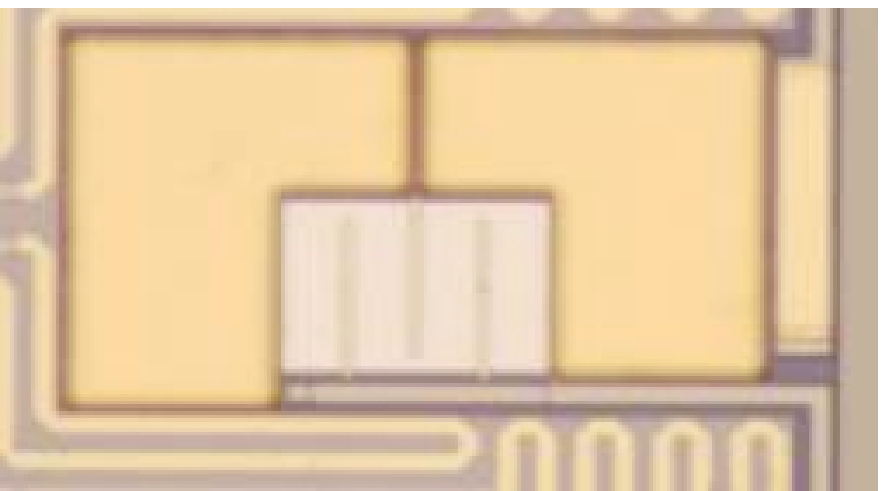}
    \end{minipage}
    &
	Symmetric Au bling 3350 nm. Usual interface. 3 Au bars each 2 $\mu$m x 43 $\mu$m x 350 nm.
    & 
    \begin{minipage}{5cm}
      24 \\ 
      83 \\ 
      33
     \end{minipage}
          &
     \begin{minipage}{.75cm}
      \includegraphics[angle=90, width=\linewidth,]{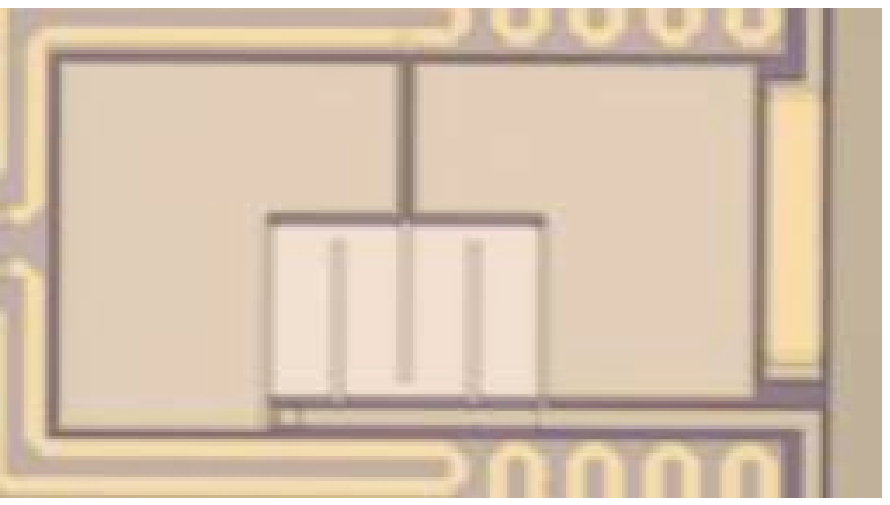}
    \end{minipage}
    &
	Symmetric PdAu bling 375nm. Usual interface. 3 PdAu bars each 2 $\mu$m x 43 $\mu$m x 375 nm.
    & 
    \begin{minipage}{5cm}
      11 \\ 
      84 \\ 
      12
    \end{minipage}
    &
     \begin{minipage}{.75cm}
      \includegraphics[angle=90, width=\linewidth,]{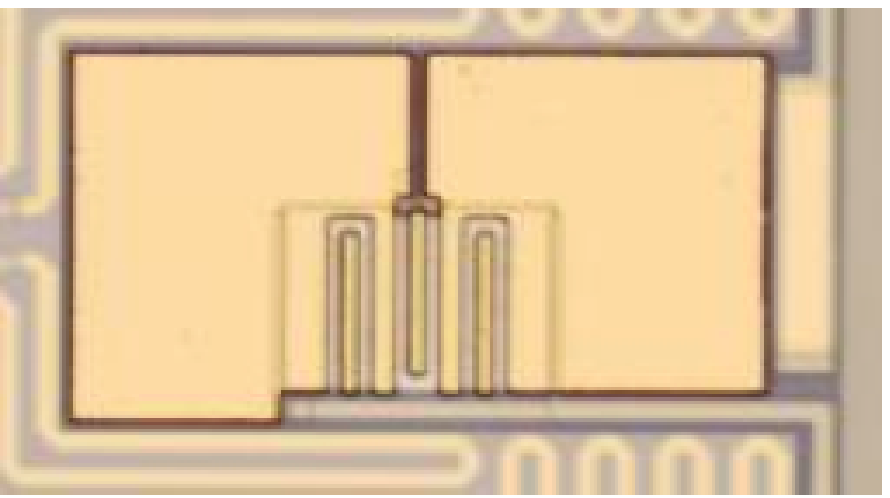}
    \end{minipage}
    &
	Symmetric Au bling 3350 nm. Usual interfaces. 3 Au bars 2 $\mu$m x 43 $\mu$m x 350 nm + 4 $\mu$m x 44 $\mu$m x 3000 nm. Au cap 3000 nm.
    & 
    \begin{minipage}{5cm}
      22 \\ 
      131 \\ 
      8 
    \end{minipage}
       &
      \\ [1.25cm] \cline{1-13}
    2
    &
    \begin{minipage}{.75cm}
      \includegraphics[angle=90,width=\linewidth,]{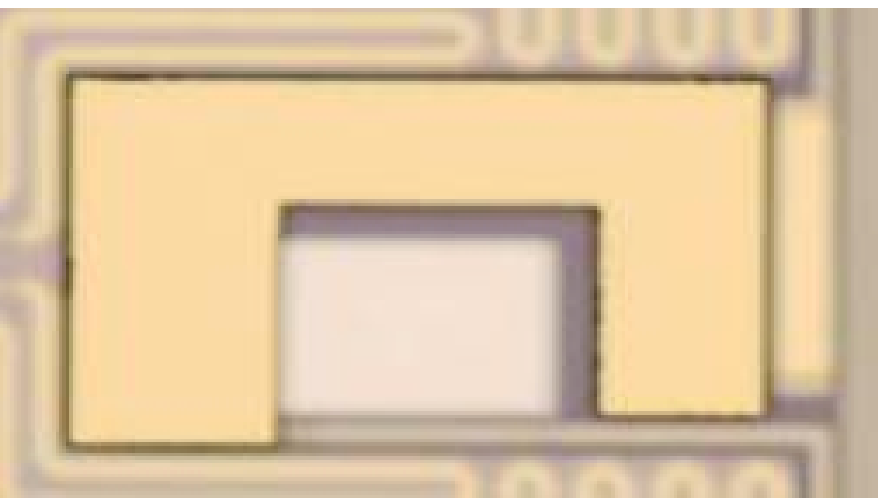}
    \end{minipage}
    &
	Monolithic Au bling 3350 nm.
    & 
    \begin{minipage}{5cm}
      17 \\ 
      52 \\ 
      30
    \end{minipage}
    &
     \begin{minipage}{.75cm}
      \includegraphics[angle=90, width=\linewidth,]{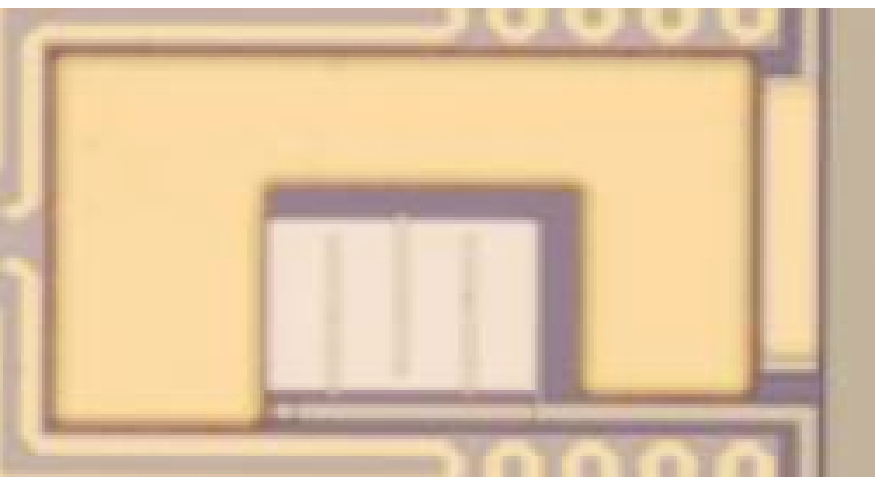}
    \end{minipage}
    &
	Monolithic Au bling 3350 nm. 3 Au bars each 2 $\mu$m x 43 $\mu$m x 350 nm.
    & 
    \begin{minipage}{5cm}
      19 \\ 
      60 \\ 
      22
     \end{minipage}
          &
     \begin{minipage}{.75cm}
      \includegraphics[angle=90, width=\linewidth,]{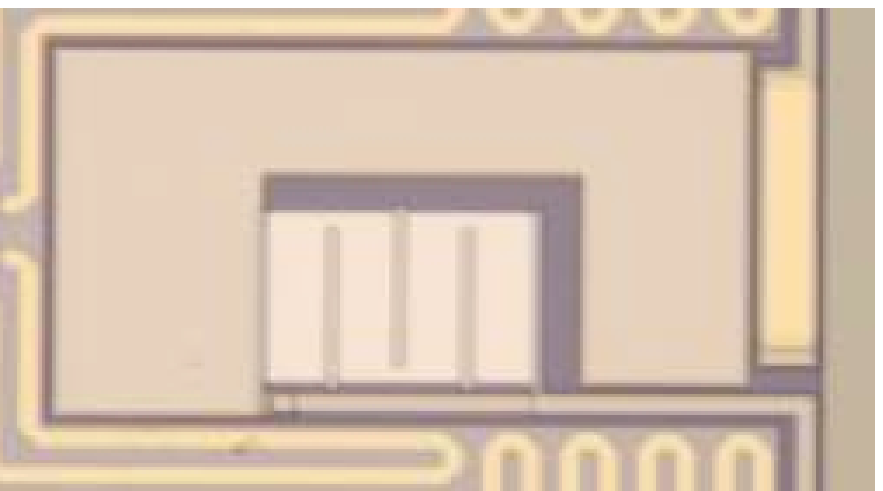}
    \end{minipage}
    &
	Monolithic PdAu bling 375 nm. 3 PdAu bars each 2 $\mu$m x 43 $\mu$m x 375 nm.
    & 
    \begin{minipage}{5cm}
      13 \\ 
      70 \\ 
      18
    \end{minipage}
    &
     \begin{minipage}{.75cm}
      \includegraphics[angle=90, width=\linewidth,]{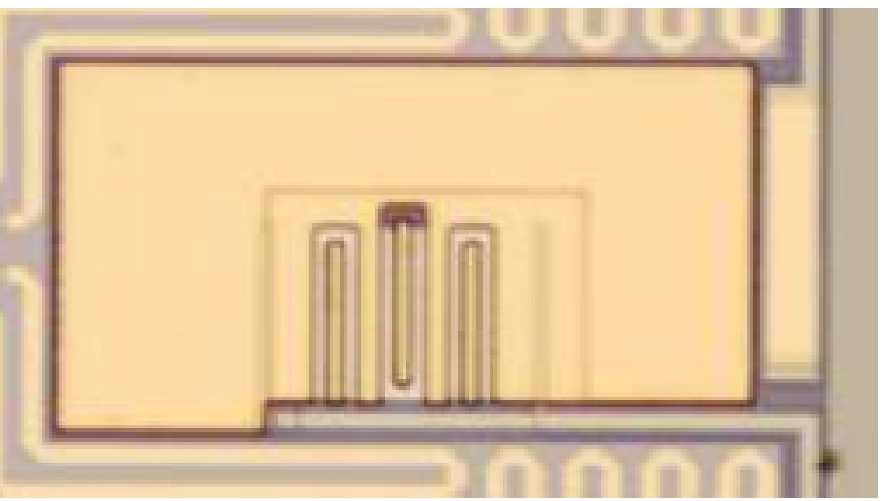}
    \end{minipage}
    &
	Monolithic Au bling 3350 nm. 3 bars 2 $\mu$m x 43 $\mu$m x 350 nm + 4 $\mu$m x 44 $\mu$m x 3000 nm. Au cap 3000 nm.
    & 
    \begin{minipage}{5cm}
      25 \\ 
      138 \\ 
      12
    \end{minipage}
       &
      \\ [1.25cm] \cline{1-13}
    3
    &
    \begin{minipage}{.75cm}
      \includegraphics[angle=90,width=\linewidth,]{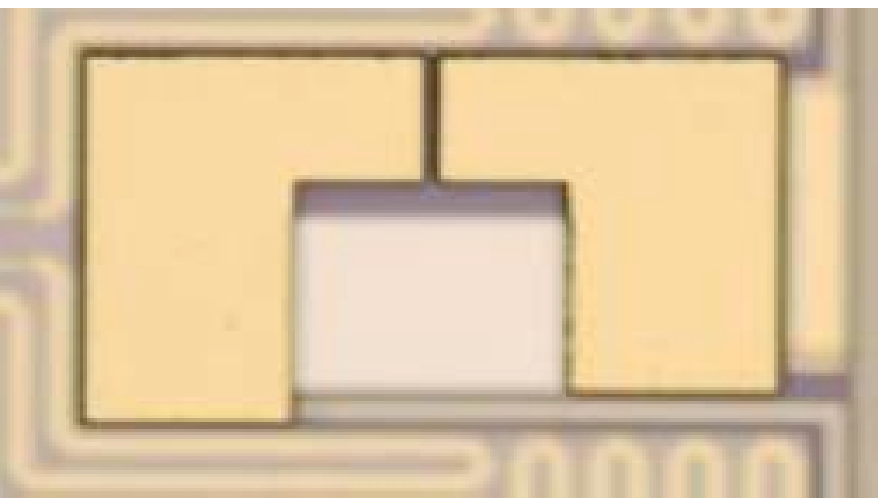}
    \end{minipage}
    &
	Symmetric Au bling 3350 nm. Extra wide LSN gap.
    & 
    \begin{minipage}{5cm}
      24\\ 
      99 \\ 
      44
    \end{minipage}
    &
     \begin{minipage}{.75cm}
      \includegraphics[angle=90, width=\linewidth,]{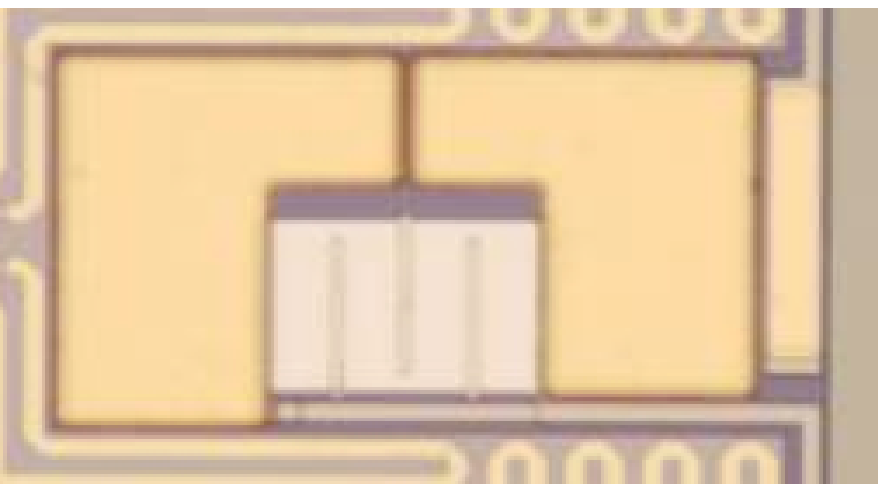}
    \end{minipage}
    &
	Symmetric PdAu  bling 375nm. Extra wide LSN gap. 3 bars each 2 $\mu$m x 43 $\mu$m x 350 nm.
    & 
    \begin{minipage}{5cm}
      21 \\ 
      73 \\ 
      29
     \end{minipage}
          &
     \begin{minipage}{.75cm}
      \includegraphics[angle=90, width=\linewidth,]{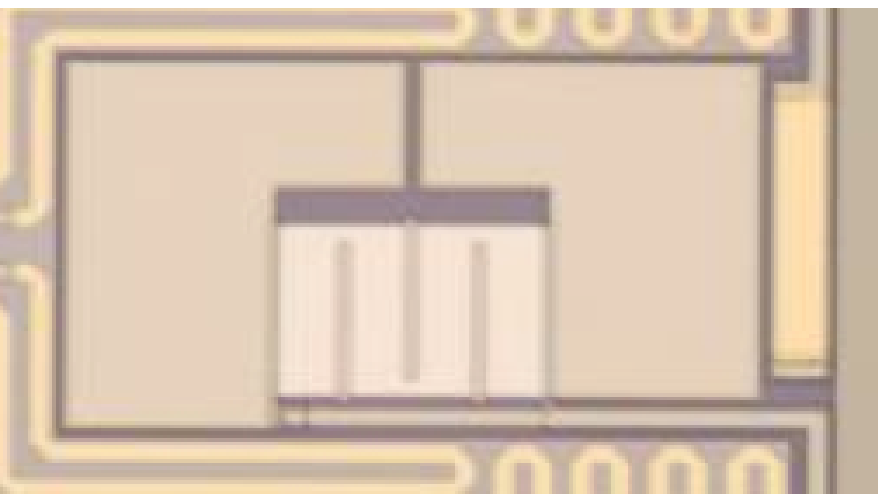}
    \end{minipage}
    &
	Symmetric PdAu bling 375nm. Extra wide LSN gap. 3 PdAu bars each 2 $\mu$m x 43 $\mu$m x 375 nm.
    & 
    \begin{minipage}{5cm}
      16 \\ 
      105 \\ 
      6
    \end{minipage}
    &
     \begin{minipage}{.75cm}
      \includegraphics[angle=90, width=\linewidth,]{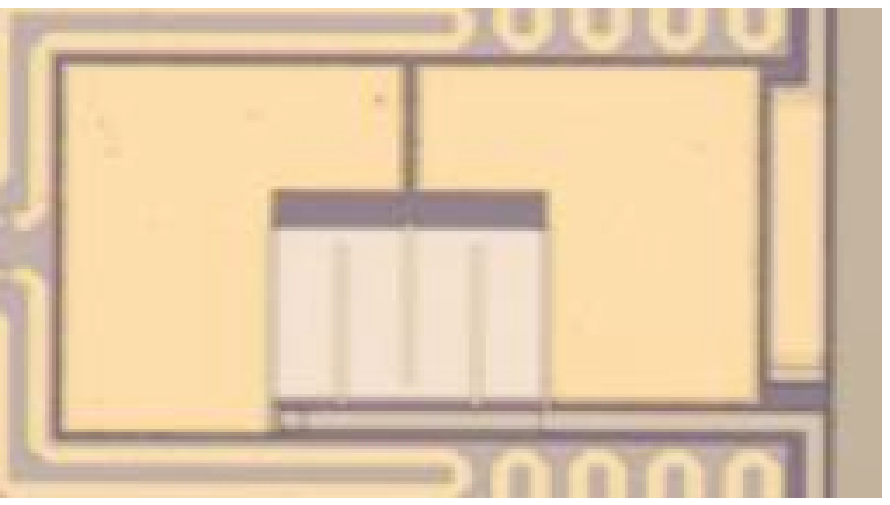}
    \end{minipage}
    &
	Symmetric Au bling 350nm. Extra wide LSN gap. 3 thin bars 2 $\mu$m x 43 $\mu$m x 350 nm.
    & 
    \begin{minipage}{5cm}
      9 \\ 
      84 \\ 
      19
    \end{minipage}
       &
      \\ [1.25cm] \cline{1-13}
    4
    &
    \begin{minipage}{.75cm}
      \includegraphics[angle=90,width=\linewidth,]{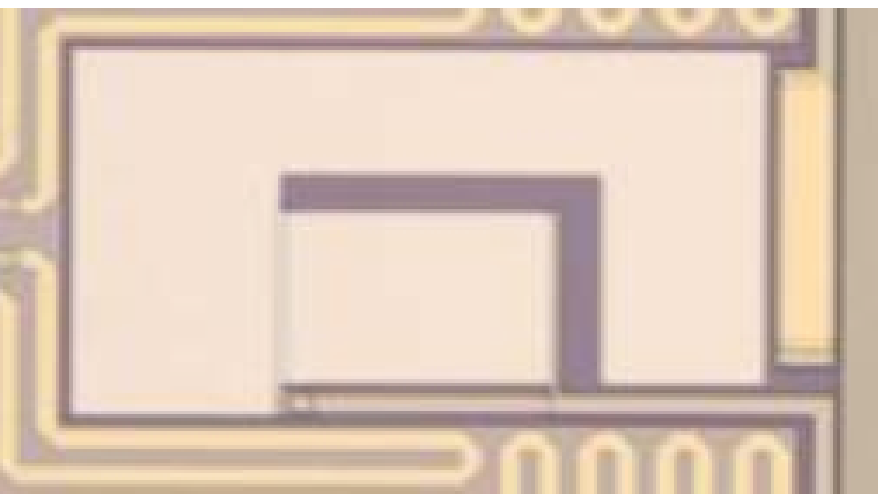}
    \end{minipage}
    &
	Monolithic AlMn bling 45 nm. 
    & 
    \begin{minipage}{5cm}
      7 \\ 
      53 \\ 
      27
    \end{minipage}
    &
     \begin{minipage}{.75cm}
      \includegraphics[angle=90, width=\linewidth, ]{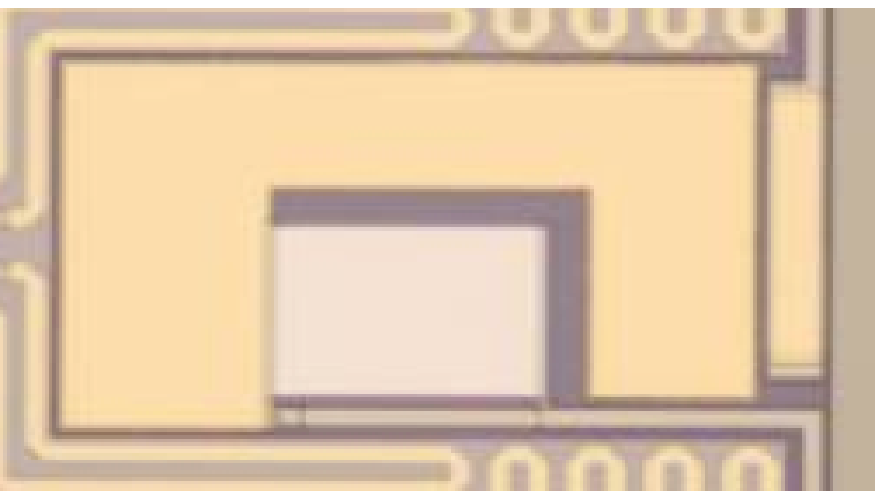}
    \end{minipage}
    &
	Monolithic Au bling 350 nm.
    & 
    \begin{minipage}{5cm}
      8 \\ 
      67 \\ 
      51
     \end{minipage}
          &
     \begin{minipage}{.75cm}
      \includegraphics[angle=90, width=\linewidth,]{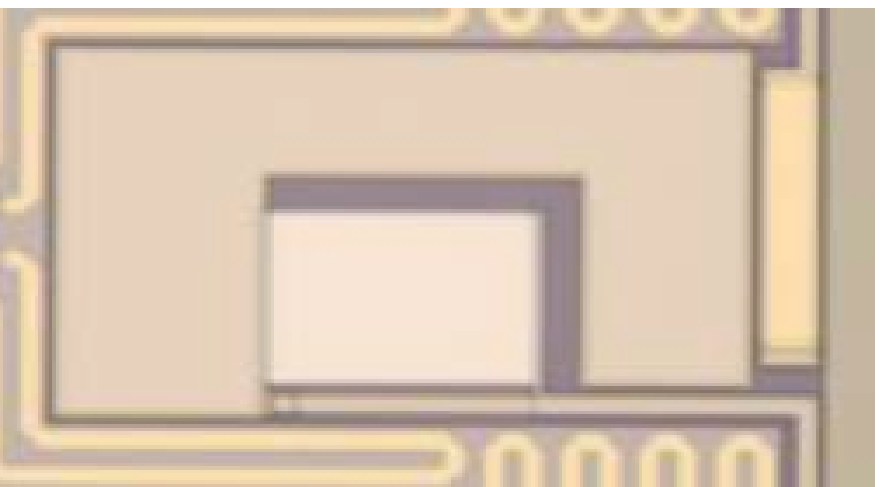}
    \end{minipage}
    &
	Monolithic PdAu bling 375 nm.
    & 
    \begin{minipage}{5cm}
      12\\ 
      62 \\ 
      25
    \end{minipage}
    &
     \begin{minipage}{.75cm}
      \includegraphics[angle=90, width=\linewidth,]{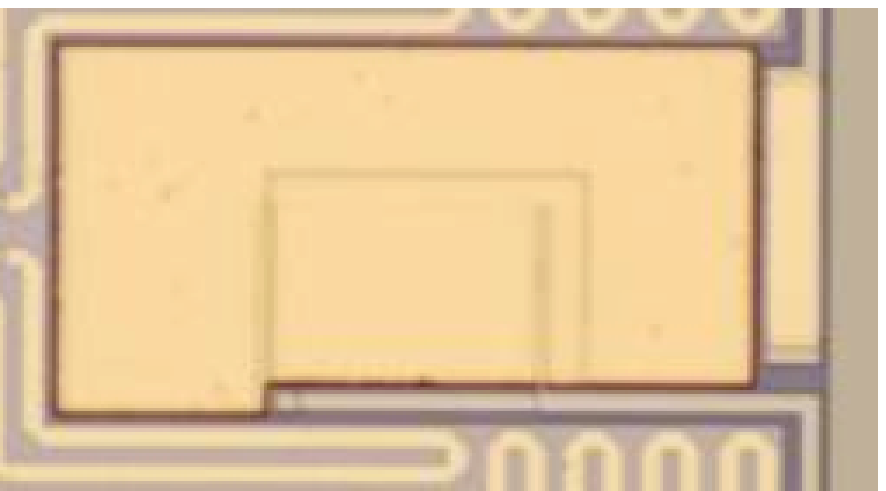}
    \end{minipage}
    &
	Monolithic Au bling 3350 nm. Au cap 3000 nm.
    & 
    \begin{minipage}{5cm}
      22 \\ 
      114 \\ 
      11
    \end{minipage}
       &
      \\ [1.25cm] \cline{1-13}
    5
    &
    \begin{minipage}{.75cm}
      \includegraphics[angle=90,width=\linewidth,]{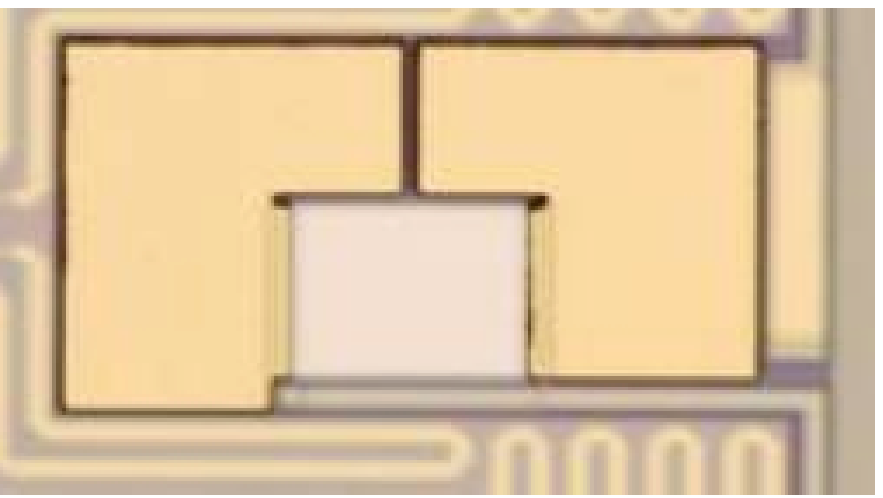}
    \end{minipage}
    &
	Symmetric Au bling 3350 nm, extends past Nb leads into TES region. 
    & 
    \begin{minipage}{5cm}
      21 \\ 
      152 \\ 
      46
    \end{minipage}
    &
     \begin{minipage}{.75cm}
      \includegraphics[angle=90, width=\linewidth, ]{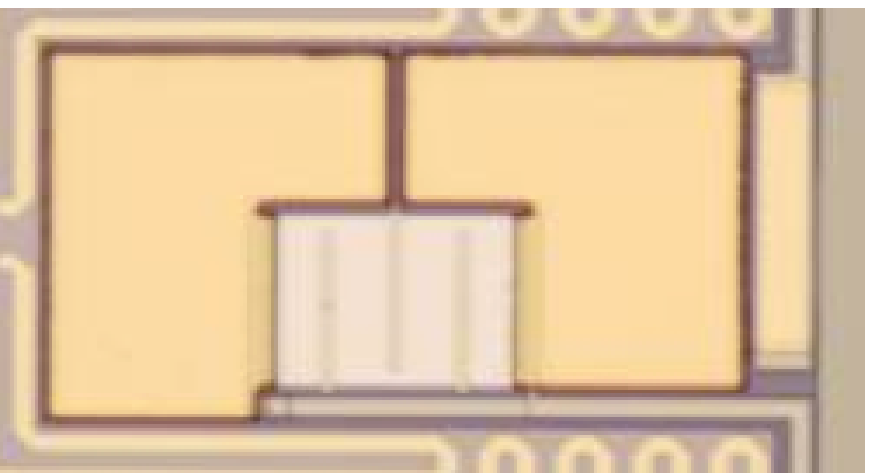}
    \end{minipage}
    &
	Symmetric Au bling 3350 nm, extends past Nb leads into TES region. 3 bars, each 2 $\mu$m x 43 $\mu$m x 350 nm.
    & 
    \begin{minipage}{5cm}
      23 \\ 
      181 \\ 
      44
     \end{minipage}
          &
     \begin{minipage}{.75cm}
      \includegraphics[angle=90, width=\linewidth, ]{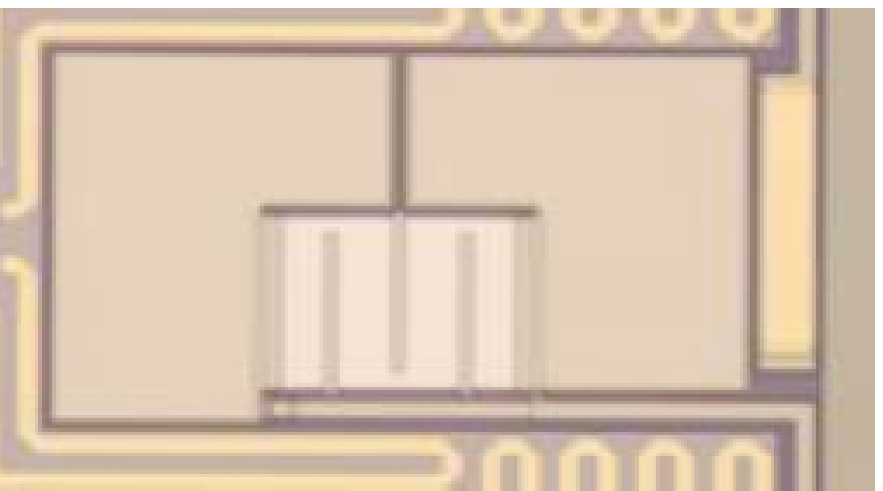}
    \end{minipage}
    &
	Symmetric PdAu bling 375 nm, extends past Nb leads into TES region. 3 PdAu bars each 2 $\mu$m x 43 $\mu$m x 375 nm.
    & 
    \begin{minipage}{5cm}
      11 \\ 
      133 \\ 
      6
    \end{minipage}
    &
     \begin{minipage}{.75cm}
      \includegraphics[angle=90, width=\linewidth,]{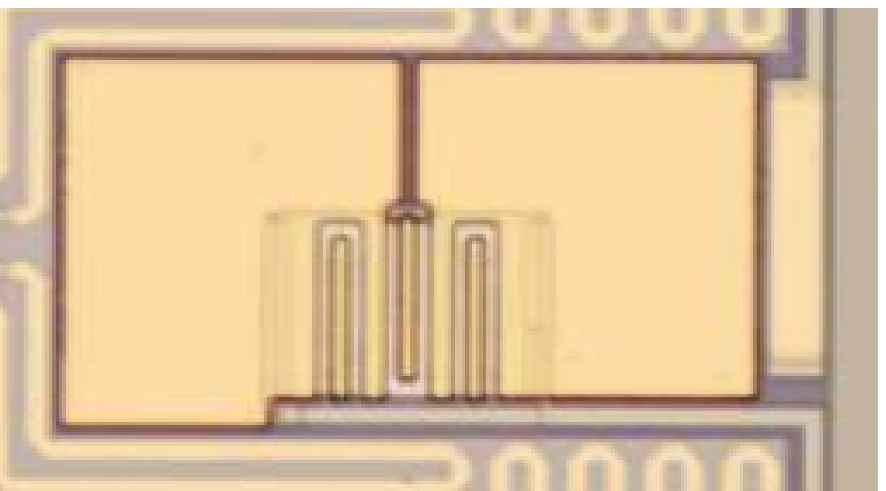}
    \end{minipage}
    &
	Symmetric Au bling 3350 nm, extends past Nb leads into TES region. 3 bars 2 $\mu$m x 43 $\mu$m x 350 nm + 4 $\mu$m x 44 $\mu$m x 3000 nm. Au cap 3000 nm.
    & 
    \begin{minipage}{5cm}
      25 \\ 
      148 \\ 
      16
    \end{minipage}
       &
      \\ [1.25cm] \cline{1-13}
    6
    &
    \begin{minipage}{.75cm}
      \includegraphics[angle=90,width=\linewidth,]{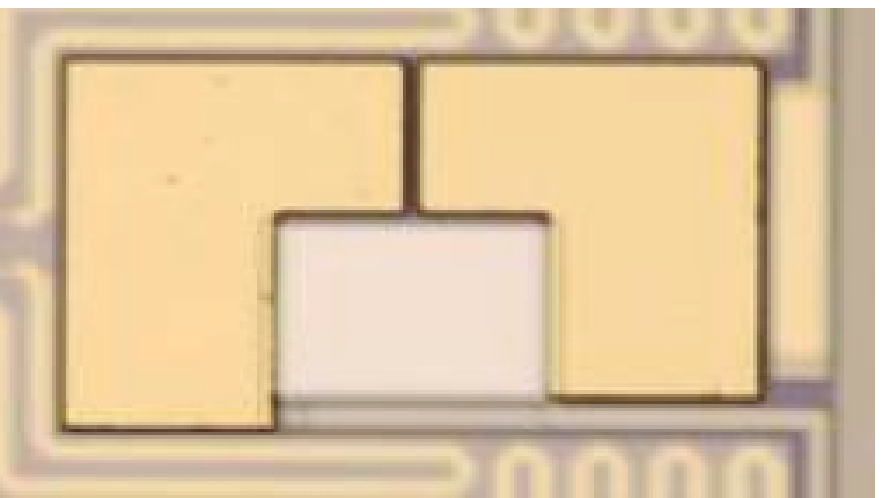}
    \end{minipage}
    &
	Symmetric Au bling 3350 nm, no AlMn under bling.
    & 
    \begin{minipage}{5cm}
      22 \\ 
      81 \\ 
      48
    \end{minipage}
    &
     \begin{minipage}{.75cm}
      \includegraphics[angle=270, width=\linewidth,]{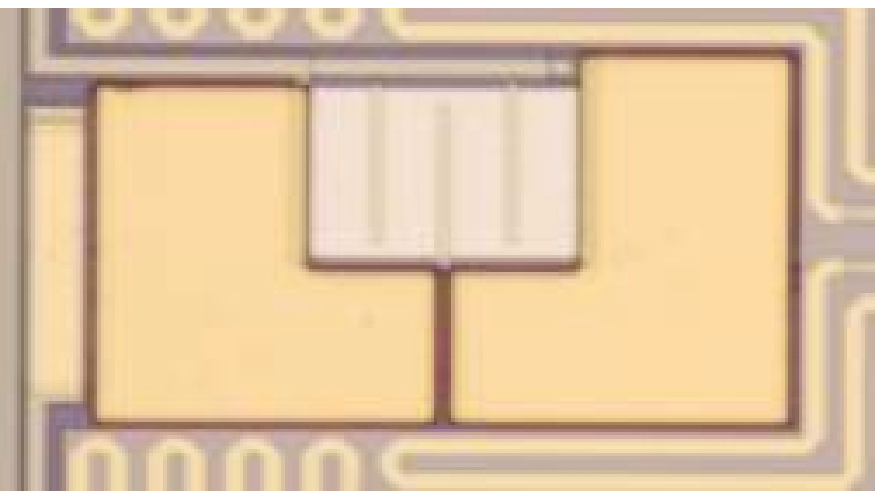}
    \end{minipage}
    &
	Symmetric Au bling 3350 nm, no AlMn under bling. 3 Au bars each 2 $\mu$m x 43 $\mu$m x 350 nm.
    & 
    \begin{minipage}{5cm}
      24 \\ 
      70 \\ 
      19
     \end{minipage}
          &
     \begin{minipage}{.75cm}
      \includegraphics[angle=270, width=\linewidth,]{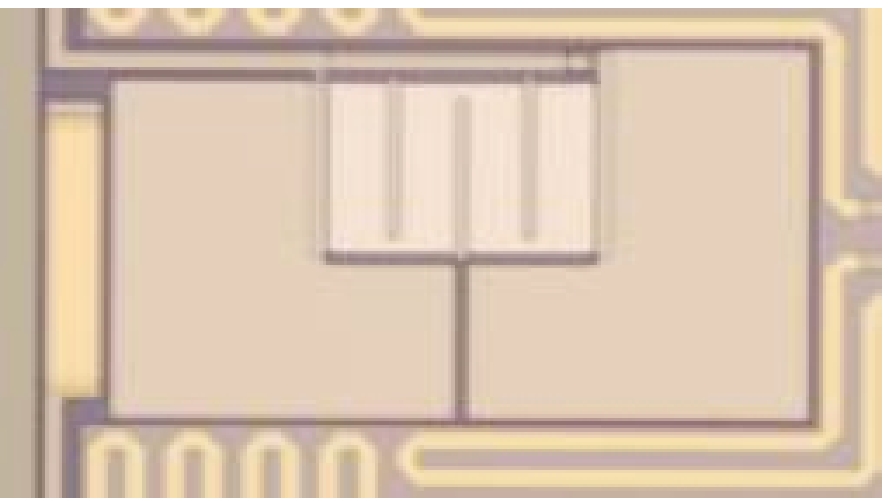}
    \end{minipage}
    &
	Symmetric PdAu bling 375 nm, no AlMn under bling. 3 PdAu bars each 2 $\mu$m x 43 $\mu$m x 375 nm.
    & 
    \begin{minipage}{5cm}
      15 \\ 
      102 \\ 
      20
    \end{minipage}
    &
     \begin{minipage}{.75cm}
      \includegraphics[angle=270, width=\linewidth, ]{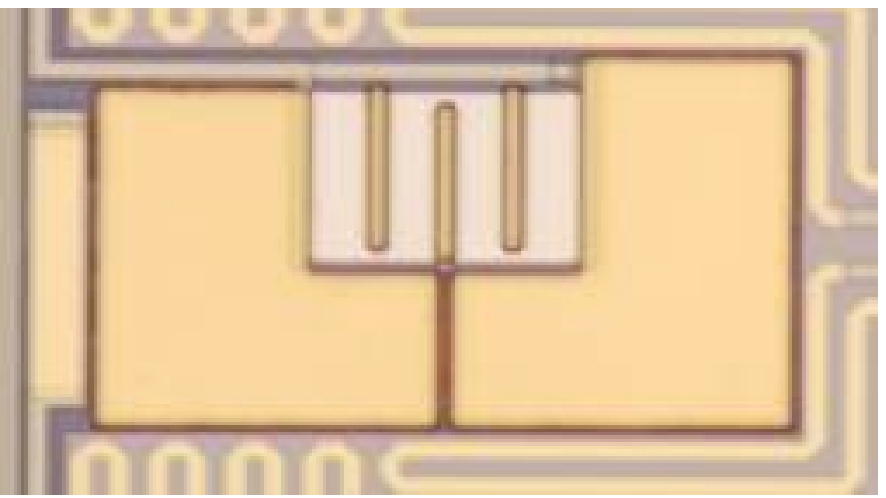}
    \end{minipage}
    &
	Symmetric Au bling 3350 nm, no AlMn under bling. 3 Au bars 2 $\mu$m x 43 $\mu$m x 350 nm + 4 $\mu$m x 44 $\mu$m x 3000 nm. 
    & 
    \begin{minipage}{5cm}
      20 \\ 
      58 \\ 
      14
    \end{minipage}
       &
      \\ [1.25cm] \cline{1-13}
    7
    &
    \begin{minipage}{.75cm}
      \includegraphics[angle=90,width=\linewidth,]{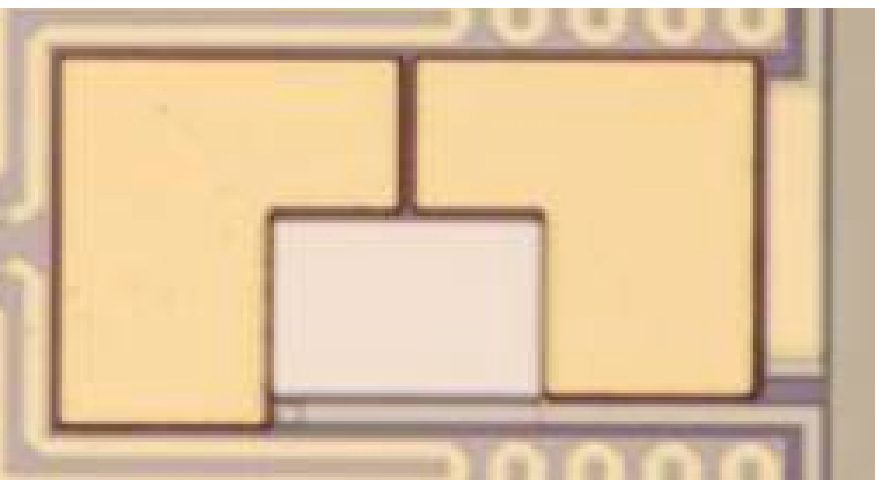}
    \end{minipage}
    &
	Symmetric Au bling 3350 nm. (Same as Type1-1)
    & 
    \begin{minipage}{5cm}
      17 \\ 
      56 \\ 
      20
    \end{minipage}
    &
     \begin{minipage}{.75cm}
      \includegraphics[angle=270, width=\linewidth, ]{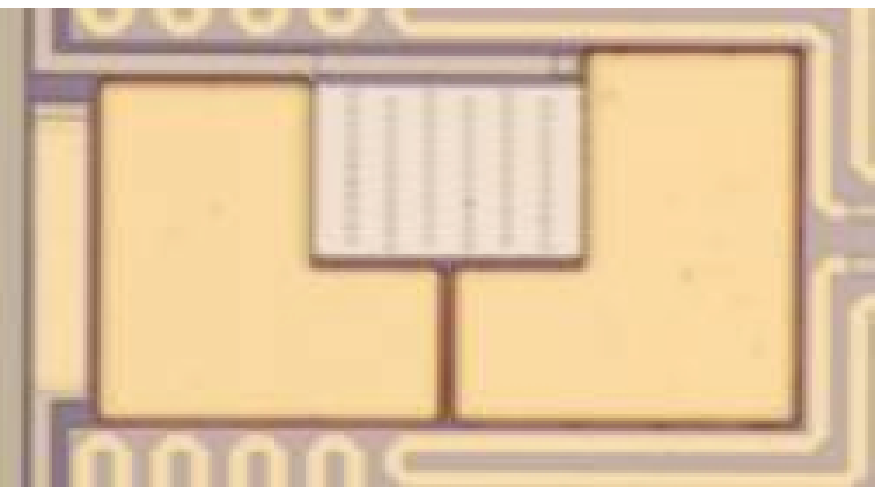}
    \end{minipage}
    &
	Symmetric Au bling 3350 nm. Au dots each 2 $\mu$m x 2 $\mu$m x 350 nm, 6 columns by 11 rows.
    & 
    \begin{minipage}{5cm}
      19 \\ 
      65 \\ 
      27
     \end{minipage}
          &
     \begin{minipage}{.75cm}
      \includegraphics[angle=270, width=\linewidth,]{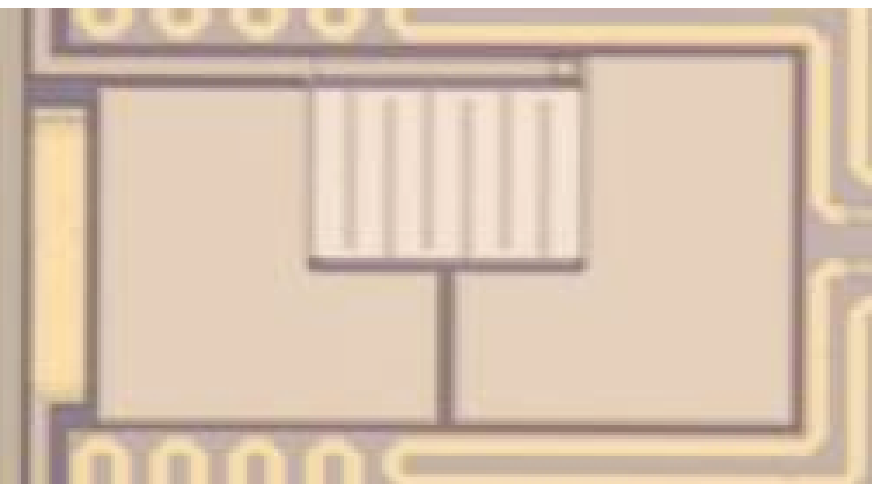}
    \end{minipage}
    &
	Symmetric PdAu bling 375 nm. PdAu dots each 2 $\mu$m x 2 $\mu$m x 375 nm, 6 columns by 11 rows.
    & 
    \begin{minipage}{5cm}
      14 \\ 
      101 \\ 
      21
    \end{minipage}
    &
     \begin{minipage}{.75cm}
      \includegraphics[angle=270, width=\linewidth]{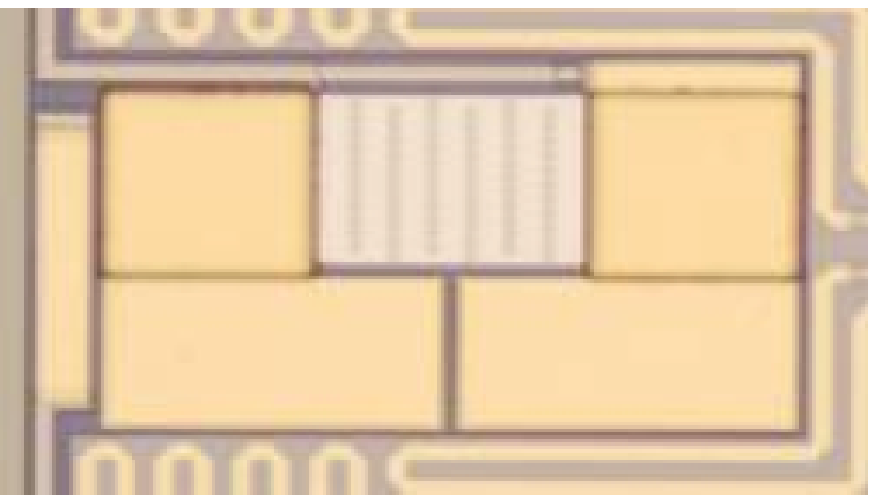}
    \end{minipage}
    &
	Au bling intermediate size 3350 nm + rest 350 nm. Au dots each 2 $\mu$m x 2 $\mu$m x 350 nm, 6 columns by 11 rows. 
    & 
    \begin{minipage}{5cm}
      18 \\ 
      85 \\ 
      17
    \end{minipage}
       &
      \\ [1.25cm] \cline{1-13}
    8
    &
    \begin{minipage}{.75cm}
      \includegraphics[angle=90,width=\linewidth]{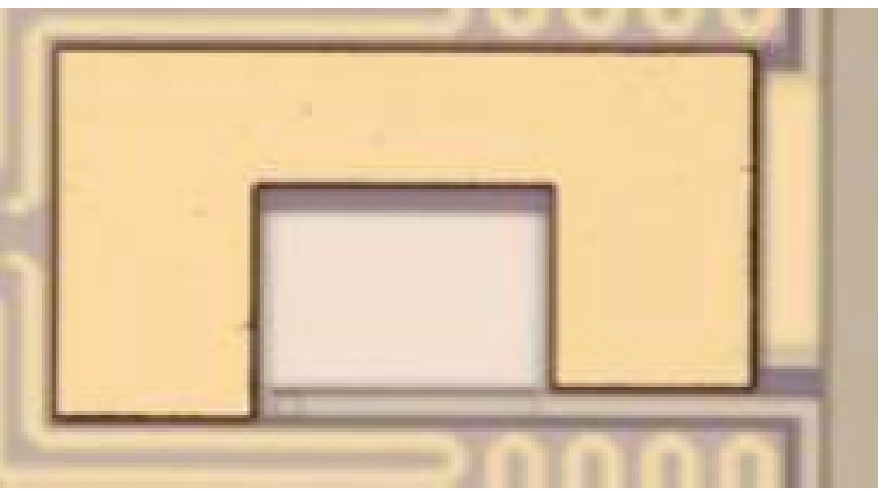}
    \end{minipage}
    &
	1 piece Au bling 3350 nm. Gap around TES. No TES/bling metal contact.
    & 
    \begin{minipage}{5cm}
      19 \\ 
      50 \\ 
      51
    \end{minipage}
    &
     \begin{minipage}{.75cm}
      \includegraphics[angle=270, width=\linewidth, ]{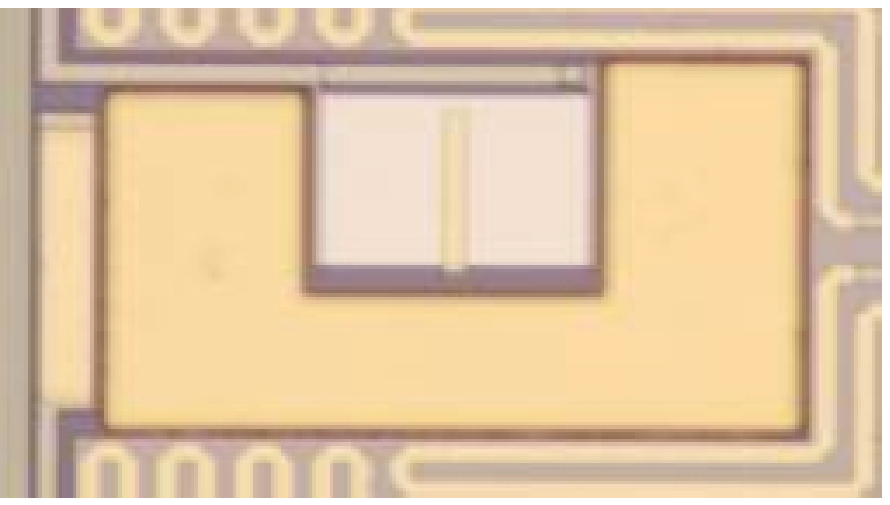}
    \end{minipage}
    &
	1 piece Au bling 3350 nm. Gap around TES. No TES/bling metal contact. 1 Au bar 5 $\mu$m x 43 $\mu$m x 350 nm.
    & 
    \begin{minipage}{5cm}
      19 \\ 
      47 \\ 
      24
     \end{minipage}
          &
     \begin{minipage}{.75cm}
      \includegraphics[angle=270, width=\linewidth, ]{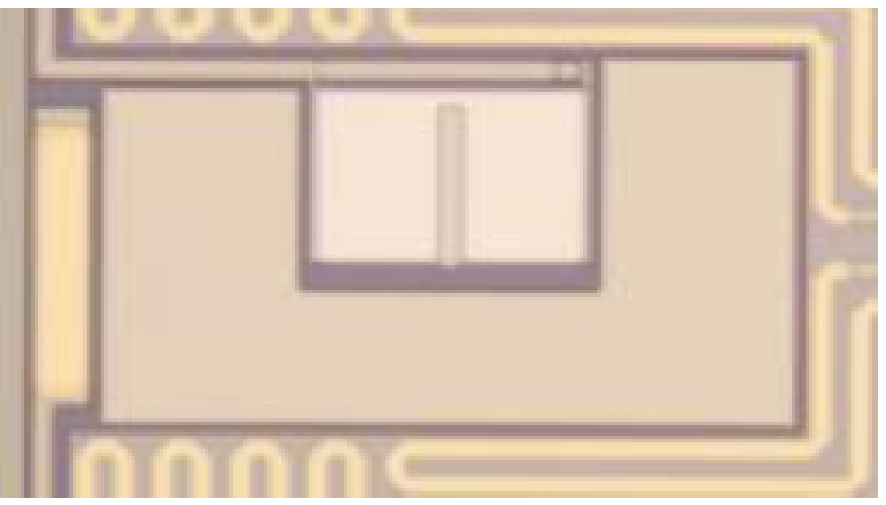}
    \end{minipage}
    &
	1 piece PdAu bling 375nm. Gap around TES. No TES/bling metal contact. 1 PdAu bar 5 $\mu$m x 43 $\mu$m x 375 nm.
    & 
    \begin{minipage}{5cm}
      13 \\ 
      55 \\ 
      30
    \end{minipage}
    &
     \begin{minipage}{.75cm}
      \includegraphics[angle=270, width=\linewidth, ]{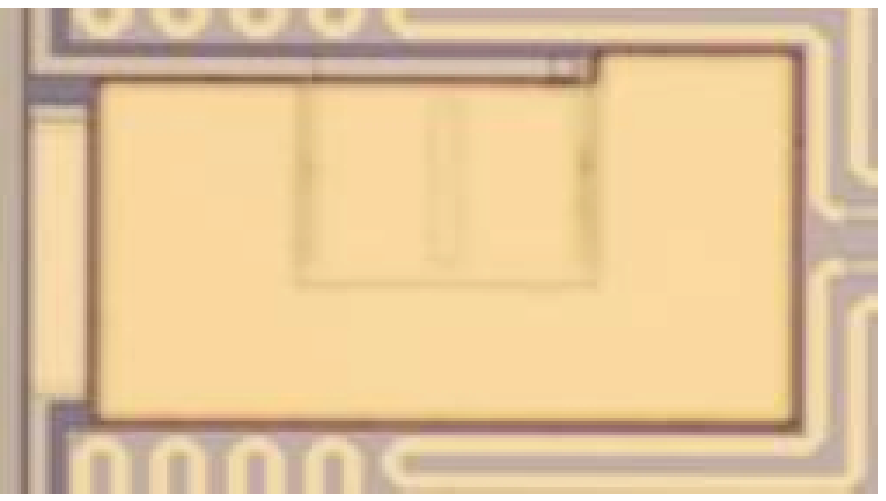}
    \end{minipage}
    &
	1 piece Au bling 3350 nm. Gap around TES. No TES/bling metal contact. 1 Au bar 5 $\mu$m x 43 $\mu$m x 3350 nm. Au cap 3000 nm.
    & 
    \begin{minipage}{5cm}
      22 \\ 
      115 \\ 
      12
    \end{minipage}
       &
      \\ [1.25cm] \cline{1-13}
    9
    &
    \begin{minipage}{.75cm}
      \includegraphics[angle=90,width=\linewidth]{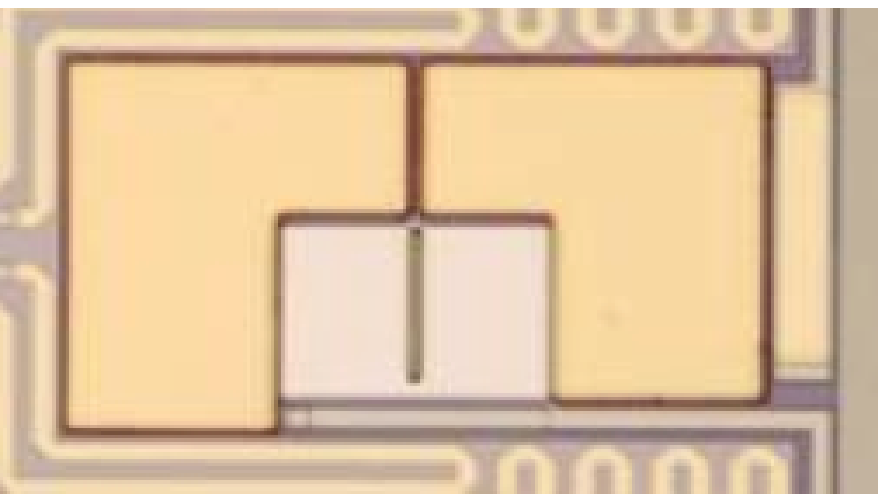}
    \end{minipage}
    &
	Symmetric Au bling 3350 nm. 1 Au bar 2 $\mu$m x 43 $\mu$m x 3350 nm.  
    & 
    \begin{minipage}{5cm}
      23 \\ 
      81 \\ 
      30
    \end{minipage}
    &
     \begin{minipage}{.75cm}
      \includegraphics[angle=270, width=\linewidth, ]{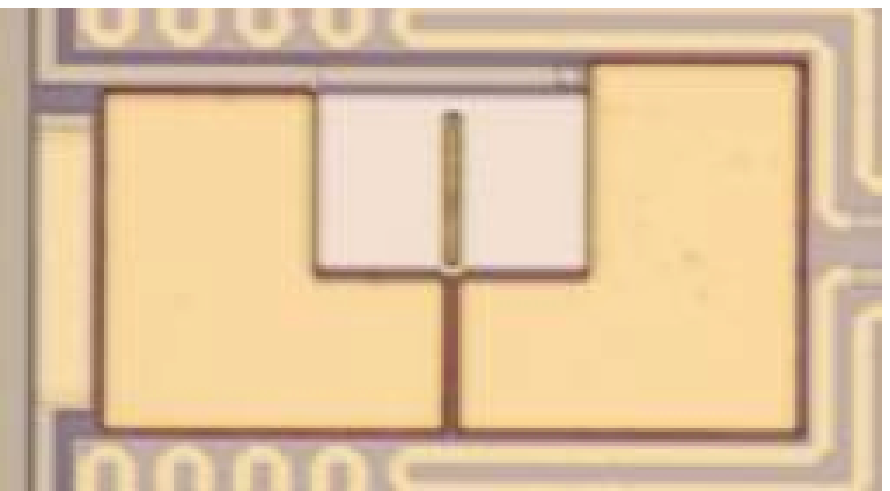}
    \end{minipage}
    &
	Symmetric Au bling 3350 nm. 1Au bar 5 $\mu$m x 43 $\mu$m x 3350 nm. Gap around TES. No TES/bling metal contact.
    & 
    \begin{minipage}{5cm}
      21 \\ 
      65 \\ 
      27
     \end{minipage}
          &
     \begin{minipage}{.75cm}
      \includegraphics[angle=270, width=\linewidth, ]{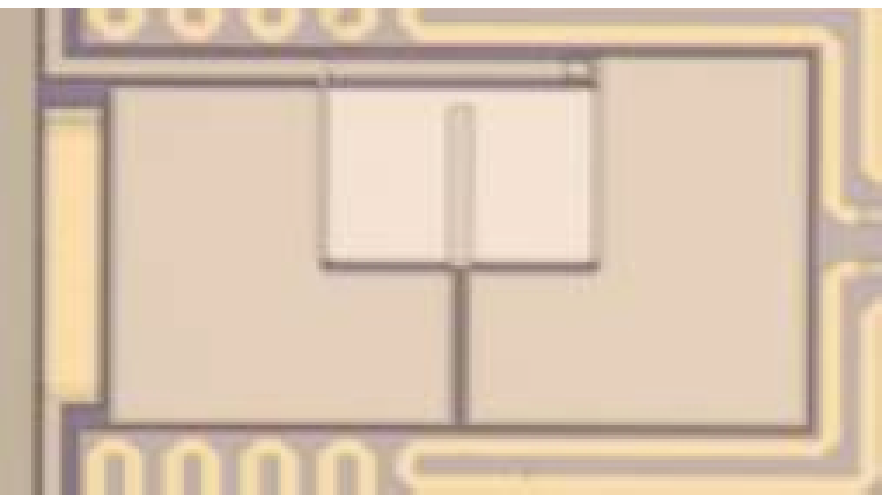}
    \end{minipage}
    &
	Symmetric PdAu bling 375 nm. 1 PdAu bar 5 $\mu$m x 43 $\mu$m x 375nm. Gap around TES. No TES/bling metal contact.
    & 
    \begin{minipage}{5cm}
      13 \\ 
      92 \\ 
      20
    \end{minipage}
    &
     \begin{minipage}{.75cm}
      \includegraphics[angle=270, width=\linewidth, ]{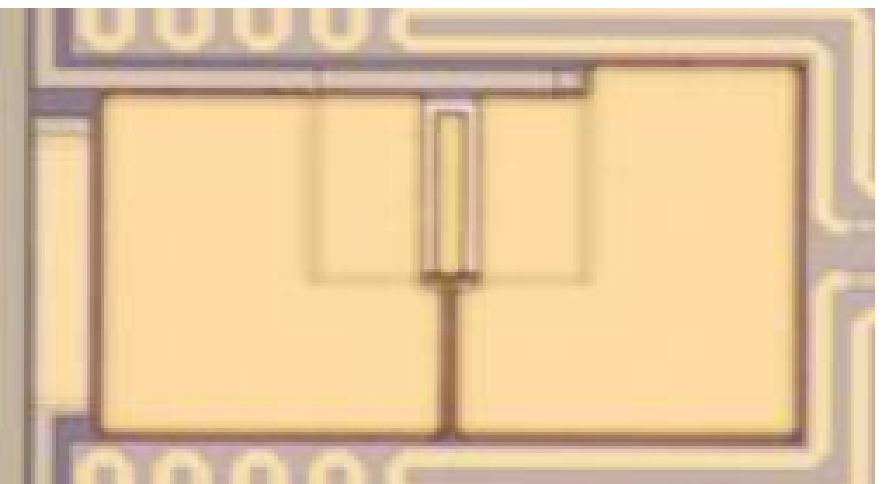}
    \end{minipage}
    &
	Symmetric Au bling 3350 nm. 1 Au bar 5 $\mu$m x 43 $\mu$m + 350 nm + 7 $\mu$m x 44 $\mu$m x 3350 nm. Au cap 3000 nm.
    & 
    \begin{minipage}{5cm}
      24 \\ 
      82 \\ 
      7
    \end{minipage}
       &
      \\ [1.25cm] \cline{1-13}
    10
    &
    \begin{minipage}{.75cm}
      \includegraphics[angle=90,width=\linewidth]{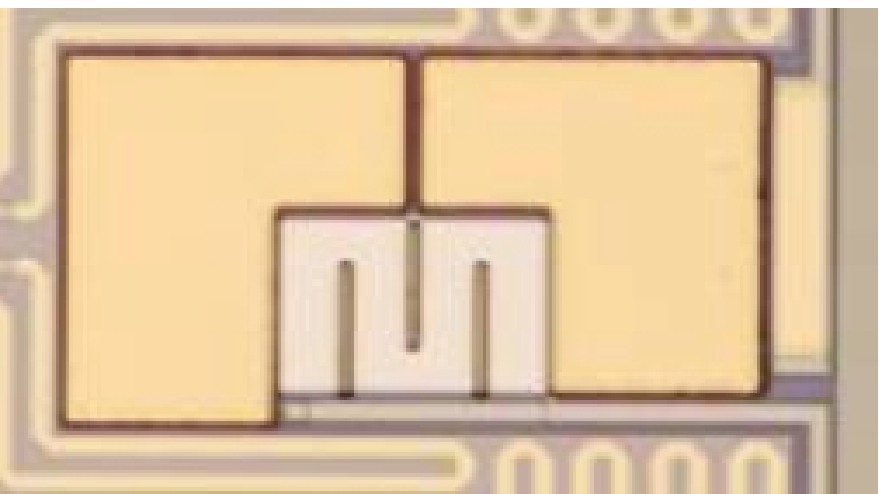}
    \end{minipage}
    &
	Symmetric Au bling 3350 nm. 3 Au bars 2 $\mu$m x 36 $\mu$m x 350 nm. 
    & 
    \begin{minipage}{5cm}
      NA \\ 
       \\ 
      
    \end{minipage}
    &
     \begin{minipage}{.75cm}
      \includegraphics[angle=270, width=\linewidth, ]{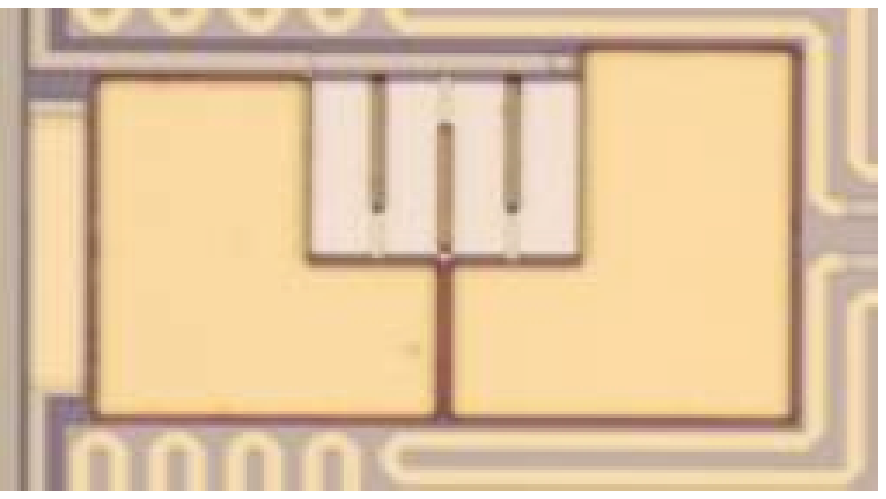}
    \end{minipage}
    &
	Symmetric Au bling 3350 nm. 3 Au bars 2 $\mu$m x 48 $\mu$m x 350 nm +75\% fractional width 2 $\mu$m x 36 $\mu$m x 350 nm. 
    & 
    \begin{minipage}{5cm}
      20 \\ 
      64 \\ 
      17
     \end{minipage}
          &
     \begin{minipage}{.75cm}
      \includegraphics[angle=270, width=\linewidth, ]{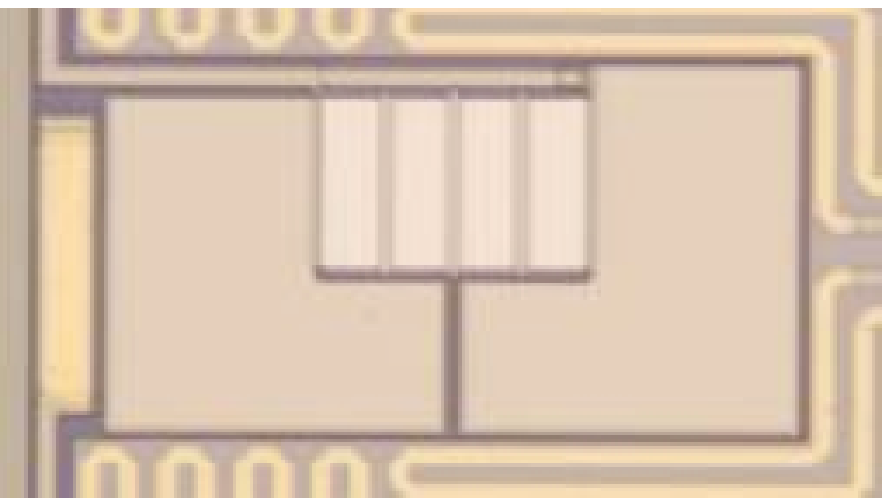}
    \end{minipage}
    &
	Symmetric PdAu bling  375 nm. 3 PdAu bars 2 $\mu$m x 48 $\mu$m x 350 nm.
    & 
    \begin{minipage}{5cm}
      12 \\ 
      87 \\ 
      16
    \end{minipage}
    &
     \begin{minipage}{.75cm}
      \includegraphics[angle=270, width=\linewidth, ]{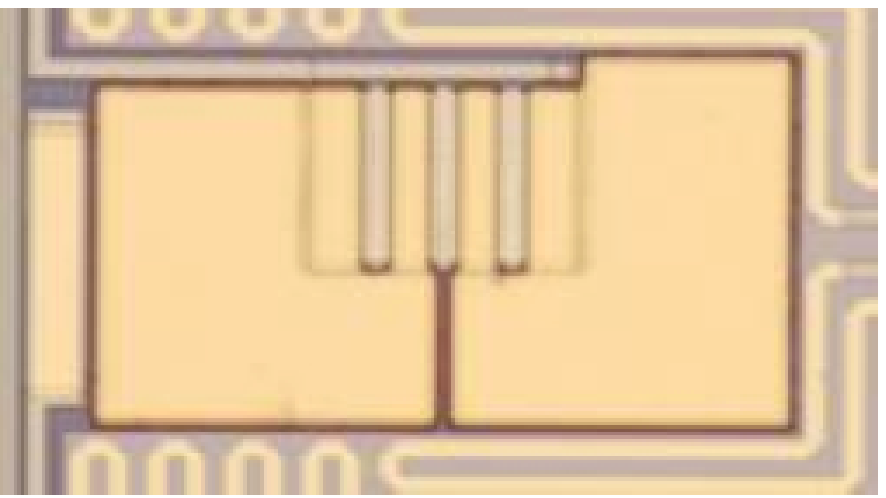}
    \end{minipage}
    &
	Symmetric Au bling 3350 nm. 3 Au bars 2 $\mu$m x 48 $\mu$m x 350 nm. Au cap 3000 nm. 
    & 
    \begin{minipage}{5cm}
      26 \\ 
      142 \\ 
      21
    \end{minipage}
       &
      \\ [1.25cm] \cline{1-13}
  \end{tabular}
  \caption{(Color online.) TES Parameters from fit to two-body thermal detector model. The intrinsic thermal time constant, $\tau_{0}$, is reported in milliseconds. $\gamma$ is the ratio of the internal to external thermal conductances $(G_{int}/G_{0})$. $\mathcal{L^*}$ is the loop gain; due to space considerations, only $\mathcal{L}(0.6)$ is reported. In cases where measurements were made on more than one device of a particular type, the fit parameters were averaged. Unless otherwise stated, the BLING is deposited on top of a layer of Al-Mn put down during the TES deposition. "Usual interfaces" denotes the BLING extending over the TES leads. Each row has a fixed TES-BLING interface, which allows comparison between the effects of adding different structures to the TES.}
  \label{tab:TESparams}
\end{sidewaystable}

\begin{acknowledgements}
Work at the University of Colorado is supported by the NSF through grant AST-0705302. Work at NIST is supported by the NIST Innovations in Measurement Science program. The McGill authors acknowledge funding from the Natural Sciences and Engineering Research Council, Canadian Institute for Advanced Research, and Canada Research Chairs program. MD acknowledges support from an Alfred P. Sloan Research Fellowship. Work at the University of Chicago is supported by grants from the NSF (awards ANT-0638937 and PHY-0114422), the Kavli Foundation, and the Gordon and Betty Moore Foundation. Work at Argonne National Lab is supported by UChicago Argonne, LLC, Operator of Argonne National Laboratory (ÒArgonneÓ). Argonne, a U.S. Department of Energy Office of Science Laboratory, is operated under Contract No. DE-AC02-06CH11357. We acknowledge support from the Argonne Center for Nanoscale Materials.\end{acknowledgements}

\end{document}